\documentclass{spie}
\usepackage{graphicx} % Required for inserting images
\usepackage{amsmath}
\usepackage{hhline}
\usepackage{float}
\usepackage{array}
\title{Developing a Compact SWIR Imaging Spectrometer for CO2 and CH4 Retrieval Using Photonic Crystal Filters}
\author[a]{Marijn Siemons}
\author[a]{Brecht Simon}
\author[a]{Irina Malysheva}
\author[a]{Ralf Kohlhaas}
\date{July 2026}
\affil[a]{SRON, Space Research Organisation Netherlands, Leiden, The Netherlands}
\authorinfo{Further author information: (Send correspondence to R.K.)\\R.K.: E-mail: r.kohlhaas@sron.nl}

\begin{document}
\maketitle
\begin{abstract}

The need for atmospheric measurements with higher spatial and temporal resolution is driving the development of satellites and satellite constellations to complement existing flagship missions. We are developing an instrument concept based on photonic crystal filters with tailored spectral transmission for trace gas retrieval. These filters can be integrated directly with the detector module, enabling a highly compact system architecture.

In this work, we present performance simulations for methane and carbon dioxide retrieval in the 1.6 µm SWIR band for a medium-resolution global coverage mission with an approximately 250 m spatial resolution and a 150 km swath. We furthermore introduce an improved retrieval algorithm that substantially reduces retrieval bias. These results demonstrate the potential of the proposed architecture for medium-resolution global greenhouse-gas mapping, with retrieval performance comparable to state-of-the-art global mapping missions such as the planned CO2M mission at substantially finer spatial resolution. This measurement approach also inherently compresses the acquired spectral information, reducing the need for high downlink data rates. In the coming year, these filters will be fabricated by NIL Technology, followed by mechanical integration and experimental validation in a breadboard system.

\end{abstract}
\section{Introduction}
Trace gas imaging instruments for earth observation rely traditionally on instrument types such as dispersive spectrometers~\cite{boni2018sentinel} or Fourier transform spectrometers~\cite{nakajima2017fourier}. Novel approaches include Fabry-Perot filters as angular-dependent bandpass filters~\cite{Jervis2021}, multi-spectral partially sampled interferograms~\cite{Simeoni2022}, or static Fourier-transform spectrometers with low-Finesse Fabry-Perot resonators~\cite{NanoCarb2}. Conventional bandpass filters are in principle an interesting approach for trace gas imaging since they can lead to compact instrument designs. Due to the push-frame scanning for satellites in low-earth orbit, each pixel can look at a different ground element, which can therefore lead to a high system etendue and with that to a high light collection. However, for high spectral resolutions and corresponding narrow bandpass windows, the large majority of the light is reflected and not detected. In addition, focused beams can already lead to substantial broadening  of the filter response at moderate f-numbers. This in turn limits the system etendue, which together with the light rejection leads to a lower performance of such an instrument. 

At SRON, we have proposed nanostructured filters in combination with computational reconstruction as a potential alternative to current trace gas imaging instruments~\cite{siemons2023compressive, Siemons2025ICSO}. Unlike conventional spectroscopy, the objective is not to reconstruct the incident spectrum, but to directly measure the spectral information required to retrieve the atmospheric state. The core idea lies in selecting filters which fit best to a given retrieval task (for example methane or carbon dioxide) from a large library of possible filters based on nanostructured elements. As a material platform, we have identified 2D photonic crystals, which are elements where the refractive index is patterned periodically in two dimensions. When the lattice constants are of the order of the wavelength of the incident light, this leads to resonance phenomena such as Fano resonances which result in diverse transmission spectra. In the short-wave-infrared (SWIR) spectral range, silicon photonic crystals with hole patterns are an attractive solution due to their technical maturity. Such photonic crystals were taken as the baseline for the current work. 

Current satellite instruments in low-earth orbit (LEO) for CH4 and CO2 detection typically fall either in the categories of global mapping instruments with low spatial resolutions or of point source detection instruments with high spatial resolutions. As examples for global mapping instruments, Sentinel-5~\cite{Sentinel5} for CH4 detection has a spatial resolution of 7.5 km x 7.5 km with an across-track (ACT) swath of 2670 km, and the planned CO2M mission~\cite{CO2Mstatus2024} with a focus on CO2 has a spatial resolution of 2 km x 2 km with a swath of 270 km. As an example for point source detection instruments, GHGSat satellites~\cite{Jervis2021} for CH4 and CO2 detection have a ground sampling distance of 25 m with a swath of 12 km. Point source detection instruments typically rely on forward motion compensation to increase the SNR, which in combination with the small swath limits the number of targets which can be detected within one day. Between low spatial resolution global mapping satellites and point source detection instruments lies a large gap in observational capabilities. In particular, medium resolution global mapping instruments would lead to a considerable increase in the knowledge of global greenhouse gas concentrations and in the detection of emission events. 

In this work, we investigate whether tailored photonic-crystal filters can enable global CH4 and CO2 mapping at approximately 250~m spatial resolution while maintaining retrieval performance comparable to current global-mapping missions. We focus for the current study on the detection in the SWIR-1 spectral range (1590~nm - 1670~nm), where no active cooling of the telescope is required. We explain the instrument model, improved computational retrieval and filter selection, and consider three different detectors for instrument modelling. For each detector type, we optimize the instrument parameters and calculate the retrieval performance. Finally, we give information about on-going work on the manufacturing and integration of photonic crystals on a flight-qualified InGaAs detector module.

%We discuss instrument configurations with photonic crystal based cameras and their predicted retrieval performance. Finally, we give an outlook on current work on the integration of such photonic crystal filters on a flight-qualified InGaAs detector module. 

%In this proceeding article, we study the potential performance of a global coverage instrument for the detection of CH4/CO2 and of another instrument for the detection of CO. We investigate a target spatial resolution of 250 m and a swath of 200 km. At the end we give an outlook on the integration of photonic crystal filters on InGaAs detector. 

% \begin{figure}
%     \centering
%     \includegraphics[width=1\linewidth]{PC filters.jpeg}
%     \caption{Illustration of photonic crystal samples in amorphous silicon (a-Si), manufactured by SRON in previous projects. (A) Photonic crystal array
% on glass. (B) Photonic crystal array on a SiN membrane. (C) Zoom-in with a scanning electron
% microscope on the sample from (B).}
%     \label{fig:PCfilter}
% \end{figure}

% \section{Instrument Concept}

\begin{figure}[h]
    \centering
    \includegraphics[width=0.55\linewidth]{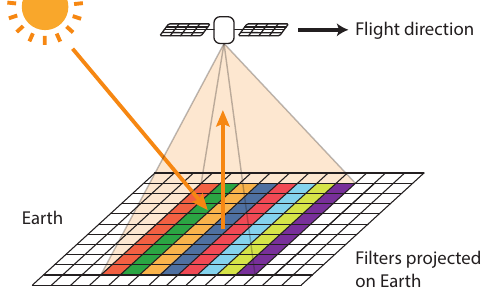}
    \caption{Illustration of the observation configuration. The photonic crystal filters map to a different position on the Earth. As the instrument flies over the Earth, the same ground resolution element is observed via different filters. Adapted from \cite{antonov2025}~.}
    \label{fig:general_concept}
\end{figure}

The instrument operates in a push-frame configuration (see Fig.~\ref{fig:general_concept}), with photonic crystals on a 2D detector. Similar to bandpass filter based instrument concepts, the instrument flies over a scene in LEO and different filters arranged as stripes on the detector see the same ground scene. The filters are specifically selected for the optimization of the trace gas retrieval. Each ground resolution element is therefore observed through all tailored filters as the spacecraft moves along-track. From the integrated signal levels per filter and detector pixels gas concentrations are retrieved by direct fitting to an atmosphere model. Since the instrument contains no dispersive elements, this leads to a very compact instrument volume. Several cameras with a telescope and a photonic crystal-detector focal plane assembly could be accommodated on a single platform in order to increase swath and precision. 

\section{Retrieval}\label{sec:retrieval}
For trace gas retrieval, the measured signal from each photonic-crystal filter is compared directly with a radiance forward model. In contrast to a conventional imaging spectrometer, the spectral response of the instrument is defined by the transmission profiles of the selected photonic-crystal filters instead of the ISRF. The atmospheric state can therefore be retrieved directly from the integrated signal levels of the different filters without reconstructing the full spectrum.

The retrieval is formulated as a nonlinear least-squares problem,
\begin{equation}
\min_{\boldsymbol{\theta}}
\sum_{k=1}^{N_\mathrm{filter}}
\left[F_k(\boldsymbol{\theta})-y_k\right]^2 ,
\end{equation}
where $F_k$ is the forward-modelled signal for filter $k$, $y_k$ is the measured signal, and $\boldsymbol{\theta}$ is the atmospheric state vector. The state vector is defined as
\begin{equation}
\boldsymbol{\theta} =
\begin{bmatrix}
\chi_\mathrm{CH4}\\
\chi_\mathrm{CO2}\\
\chi_\mathrm{H2O}\\
\alpha_0\\
\alpha_1
\end{bmatrix},
\end{equation}
where $\chi_m$ denotes the column density of gas $m$, and $a_0$ and $a_1$ describe a constant and linear component of the surface albedo, respectively. Water vapour is included as a retrieval parameter because of its absorption features in the selected SWIR spectral range. The state vector is separated into the albedo parameters $\boldsymbol{\theta}_\alpha = [\alpha_0, \alpha_1]$ and gas parameters $\boldsymbol{\theta}_\chi = [\chi_\mathrm{CH4},  \chi_\mathrm{CO2}, \chi_\mathrm{H2O}]$.

The forward model for filter $k$ is given by
\begin{equation}
F_k(\boldsymbol{\theta}) =
t_\mathrm{int} G N_\mathrm{pxl}
\int_{\lambda_\mathrm{min}}^{\lambda_\mathrm{max}}
a(\boldsymbol{\theta}_\alpha, \lambda)
T_k(\lambda)
S(\boldsymbol{\theta}_\chi,\lambda)
d\lambda ,
\end{equation}
where $t_\mathrm{int}$ is the integration time, $G$ is the optical etendue, $N_\mathrm{pxl}$ is the number of detector pixels contributing to filter $k$, $a(\boldsymbol{\theta}_\alpha, \lambda)$ the Earth's spectrally dependent albedo, $T_k$ is the photonic crystal filter transmission, and  $S(\boldsymbol{\theta}_\chi, \lambda)$ is the (non-scattering) atmosphere model which depends on the gas concentrations and is given by
\begin{equation}
S(\boldsymbol{\theta}_\chi, \lambda) = \frac{ \cos(\zeta_s) }{\pi}E(\lambda) \exp \left[-r_\mathrm{air}\sum_m \theta_{\chi_m}\sum_z  \sigma_m (\lambda, z) c_m (z) \right]
\end{equation}
with $\zeta_s$ the sun zenith angle, $E$ the sun's irradiance, $r_\mathrm{air}$ the air mass factor and $c_m(z)$ the (vertical) column number density profile at atmosphere layer $z$ and $\sigma_m$ the absorption cross-section \cite{Siemons2024}.

An important challenge was identified when applying a conventional nonlinear least-squares retrieval to the highly compressed spectral measurements. At low signal levels, the nonlinear forward model combined with measurement noise can result in curvature-induced retrieval bias. This effect is particularly relevant for the proposed instrument because the limited number of optimized spectral measurements increases the correlation between the gas concentrations and the albedo parameters.

To reduce this bias, we use variable projection \cite{GolubPereyra2003} which has been used previously in remote sensing \cite{OLeary2013}. This method separates the linear albedo parameters from the nonlinear gas parameters and solves them independently. For a fixed set of gas concentrations, the forward model can be written as
\begin{equation}
F(\boldsymbol{\theta}_\chi,\boldsymbol{\theta}_\alpha)
=
\alpha_0 A_0(\boldsymbol{\theta}_\chi)
+
\alpha_1 A_1(\boldsymbol{\theta}_\chi),
\end{equation}
where
\begin{equation}
A_0^k(\boldsymbol{\theta}_\chi)
=
t_\mathrm{int} G N_\mathrm{pxl}
\int_{\lambda_\mathrm{min}}^{\lambda_\mathrm{max}}
T_k(\lambda)
S(\boldsymbol{\theta}_\chi,\lambda)
d\lambda
\end{equation}
and
\begin{equation}
A_1^k(\boldsymbol{\theta}_\chi)
=
t_\mathrm{int} G N_\mathrm{pxl}
\int_{\lambda_\mathrm{min}}^{\lambda_\mathrm{max}}
\lambda T_k(\lambda)
S(\boldsymbol{\theta}_\chi,\lambda)
d\lambda .
\end{equation}
These terms form the design matrix
\begin{equation}
A(\boldsymbol{\theta}_\chi)
=
\left[
A_0(\boldsymbol{\theta}_\chi),
A_1(\boldsymbol{\theta}_\chi)
\right].
\end{equation}
For fixed gas concentrations, the albedo parameters can then be obtained analytically from a weighted linear least-squares solution,
\begin{equation}
\hat{\boldsymbol{\theta}}_a
=
\left(A^TWA\right)^{-1}A^TWy ,
\end{equation}
where $W$ is the weighting matrix corresponding to the assumed Poisson noise statistics. The nonlinear optimization is consequently reduced to the gas parameters only. This reduces the dimensionality of the nonlinear optimization and, more importantly, removes the need to numerically optimize the correlated albedo parameters.

The resulting reduced residual can be expressed using the reduced projection matrix
\begin{equation}
P = 
A(A^TWA)^{-1}A^TW
\end{equation}
as
\begin{equation}
r(\boldsymbol{\theta}_\chi)
=
(I-P(\boldsymbol{\theta}_\chi))y .
\end{equation}
The gas concentrations are subsequently obtained by minimizing the reduced nonlinear least-squares problem using a standard nonlinear least-squares optimizer (such as \emph{least\_squares} from \emph{scipy.optimize}). The effect of this approach on the retrieval precision and bias is evaluated in Section~\ref{sec:performance}.
\begin{figure}[H]
\centering
\includegraphics[trim={1cm, 0, 1cm, 0}, clip, width=0.9\linewidth]{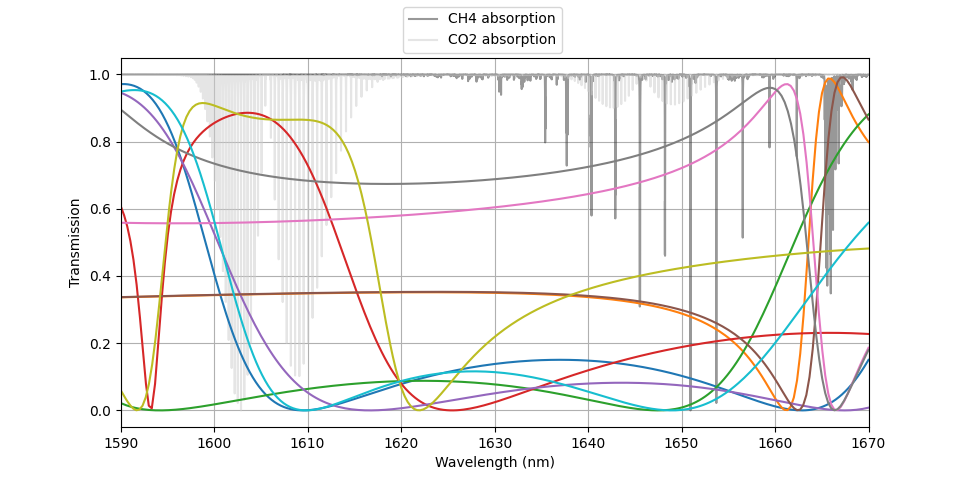}
\caption{Transmission profiles of the ten photonic-crystal filters selected for the combined CO$_2$ and CH$_4$ retrieval in the SWIR-1 band.}
\label{fig:concepta_filterset}
\end{figure}
\section{PC filter simulations and selection}
A large library of photonic crystal filters was simulated using rigorous coupled-wave analysis (RCWA) in Lumerical. The filters consist of periodic structures with four different geometries: circles, rectangles, diamonds and crosses. The pitch was varied from 600~nm to 1000~nm in steps of 25~nm, while the feature dimensions were varied between 25\% and 75\% of the pitch in steps of 5~nm. The maximum pitch of 1~\textmu m was chosen to avoid the generation of diffraction orders within the spectral range of interest. In total, this resulted in a library of approximately 6500 physically realizable filter transmission profiles.

From this library, filters were selected specifically to maximize the information available for the CH$_4$ and CO$_2$ retrieval. The selection method is an extension of the approach presented in our previous work~\cite{Siemons2024}. Rather than selecting filters based on their individual spectral characteristics, combinations of filters are evaluated based on the retrieval performance of the complete filter set. This allows the filters to provide complementary information on CH$_4$ and CO$_2$, while simultaneously separating these signals from H$_2$O and the surface-albedo parameters.

The selection procedure consists of three stages. First, an information-based pre-selection is used to reduce the full filter library to the most promising candidates. Second, combinations of these filters are evaluated using the Cramér--Rao lower bound (CRLB) of the retrieval parameters. Since the CRLB can be evaluated efficiently, a very large number of candidate filter sets can be screened. Finally, the best-performing sets are evaluated using Monte Carlo retrieval simulations, thereby including the nonlinear behaviour and potential bias of the actual retrieval.

Compared with our previous work~\cite{Siemons2024}, the addition of H$_2$O and a linear albedo term increases the state vector to five parameters and consequently requires at least five independent filter measurements. Direct evaluation of all five-filter combinations from a sufficiently large pre-selection would result in a prohibitively large number of combinations. We therefore use a hierarchical selection procedure. An initial pre-selection of 700 filters is first evaluated using four-filter subsets for two reduced state vectors: $(a_0,a_1,\mathrm{CO_2},\mathrm{CH_4})$ and $(a_0,\mathrm{H_2O},\mathrm{CO_2},\mathrm{CH_4})$. The 125 best-performing filters from each evaluation are combined into a set of 250 candidate filters. Five-filter combinations are subsequently formed from this reduced library, resulting in approximately $10^{10}$ candidate sets, for which the CRLB of the complete five-parameter retrieval is evaluated.

The 1000 filter sets with the best predicted performance are subsequently evaluated using Monte Carlo retrieval simulations. Based on these results, a final set of ten complementary filters was selected for the instrument simulations presented in this work. Their transmission profiles are shown in Fig.~\ref{fig:concepta_filterset}.
 
\section{Instrument configurations}
We next assess the performance of the proposed architecture for a medium-resolution global mapping mission, targeting a spatial resolution of approximately 250~m and an across-track (ACT) swath of at least 100~km. Three instrument configurations were developed based on different detector technologies: the Teledyne CHROMA-D 3K $\times$ 512 MCT detector, the Lynred COBRA-L MCT detector, and the Lynred SNAKE InGaAs detector. The resulting configurations are summarized in Table~\ref{tab:config}.

The CHROMA-D provides a large detector format and allows a 150~km ACT swath using a single focal-plane assembly. The COBRA-L similarly provides a large MCT focal plane, but its lower maximum readout frequency requires a larger F-number to prevent detector saturation. The SNAKE detector is considerably smaller, and we therefore consider a modular configuration consisting of a $3\times3$ array of telescope and detector modules. The use of an InGaAs detector makes this configuration particularly attractive for a compact instrument, as it does not require cryogenic detector cooling. The CHROMA-D detector will be used for the CHIME mission \cite{Buschkamp2025} and the SNAKE detector for the TANGO SCOUT mission \cite{Cooney2025}.

The three configurations were sized using the same design criteria. The ground instantaneous  field of view (GiFOV) of a single pixel was chosen such that (1) an ACT swath of at least 100~km is achieved, (2) a ground resolution element (GRE) close to 250~m can be obtained using integer spatial binning, and (3) the detector full-well capacity is not exceeded for an albedo of 0.8 at a solar zenith angle of 0\textdegree. For the CHROMA-D configuration, the large number of ACT pixels allows the swath to be extended to 150~km. For the COBRA-L configuration, an asymmetric GiFOV is used to limit the ALT extent of the field of view to 30~km, ensuring that all filters observe the same ground element within approximately $\pm2$\textdegree. Furthermore, as mentioned before, the relatively low maximum readout frequency of the COBRA-L requires an F-number of 9.2 to prevent saturation.

In the along-track (ALT) direction, the motion of the spacecraft during the observation determines the spatial response of an individual measurement. Each detector sample integrates over the full ALT smear distance, while temporal coadding determines the number of detector frames acquired during this interval. Subsequent spatial binning combines these measurements into the final GRE. The ALT binning width therefore plays a role analogous to the slit width in a conventional push-broom spectrometer which together with the spacecraft motion determines the shape of the system energy distribution function (SEDF).

\begin{figure}[]
    \centering
    \includegraphics[width=1\linewidth]{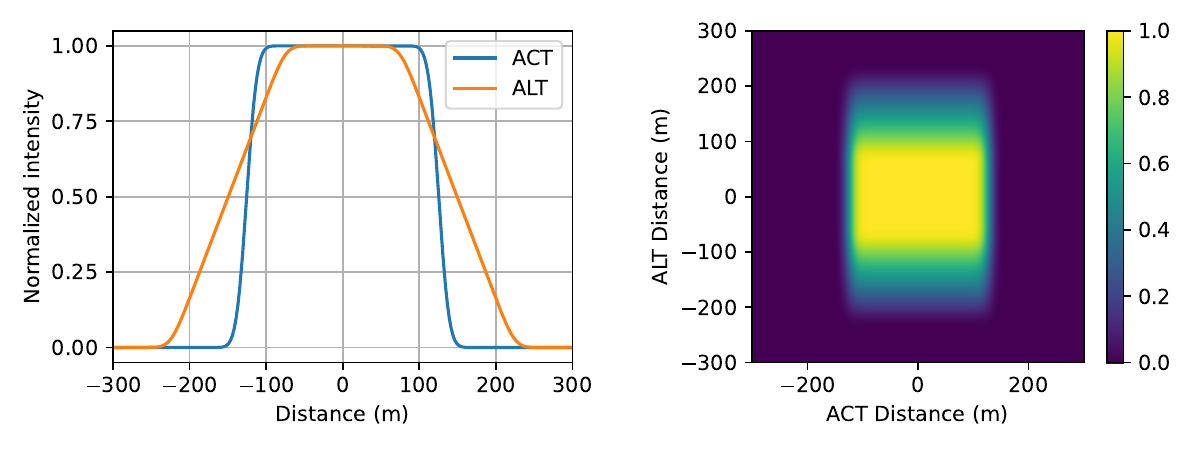}
    \caption{Example SEDF cross-section and 2D illustration for the CHROMA-D configuration.}
    \label{fig:sedf}
\end{figure}

For the CHROMA-D configuration, for example, the resulting data product is sampled at $250\times150$~m in the ACT and ALT directions, respectively. Each sample has a rectangular SEDF in the ACT direction with a width of 250~m and a trapezoidal response in the ALT direction with a FWHM of approximately 300~m and a total extent of 450~m. Figure~\ref{fig:sedf} shows the resulting SEDF, including an additional Gaussian contribution of 25~m FWHM representing the combined effect of the optical PSF and spacecraft jitter. The 150~m ALT sampling therefore oversamples the effective spatial resolution. Alternatively, increased ALT binning could be applied to improve the retrieval precision at the expense of a broader spatial response.

\begin{table}[]
\centering
 \caption{Instrument configuration for three different detectors, aimed for a medium resolution global coverage instrument.}
 \vspace{5pt}
\begin{tabular}{|l|c|c|c|l|} \hline
                          \textbf{Parameter} & \textbf{CHROMA-D} & \textbf{COBRA-L} & \textbf{SNAKE 3x3} & \textbf{unit} \\  \hline  \hline
Swath ACT                 & 150               & 100              & 115                & km            \\ \hline
Swath ALT                 & 25                & 30               & 30                 & km            \\ \hline
GRE                       & 250 x 300         & 275 x 300        & 240 x 240          & m             \\ \hline
GiFoV (ACT x ALT)          & 50 x 50           & 55 x 30          & 60 x 60            & m             \\ \hline
Binning (ACT x ALT)         & 5 x 3             & 5 x 5            & 3 x 3              &               \\ \hline
Temp. coadding            & 8                 & 3                & 6                  &               \\ \hline
Detector format (ACT x ALT) & 3K x 512          & 1872 x 1112      & 640 x 512          &               \\ \hline
F\#                       & 4                 & 9.2              & 4                  &               \\ \hline
Focal length (ACT x ALT)    & 0.18              & 0.18 x 0.33      & 0.125         & m             \\ \hline
Aperture (ACT x ALT)        & 45 x 45           & 20 x 36          & 31 x 31            & mm            \\ \hline
Readout freq.              & 225               & 78               & 265                & Hz            \\ \hline
Full well capacity        & 2.6               & 1.3              & 1.6                & Me$^-$            \\ \hline
Pixel size                & 18                & 20               & 15                 & um            \\ \hline
       
\end{tabular}
\label{tab:config}
\end{table}

\section{Performance} \label{sec:performance}
\begin{figure}
    \centering
    \includegraphics[width=\linewidth]{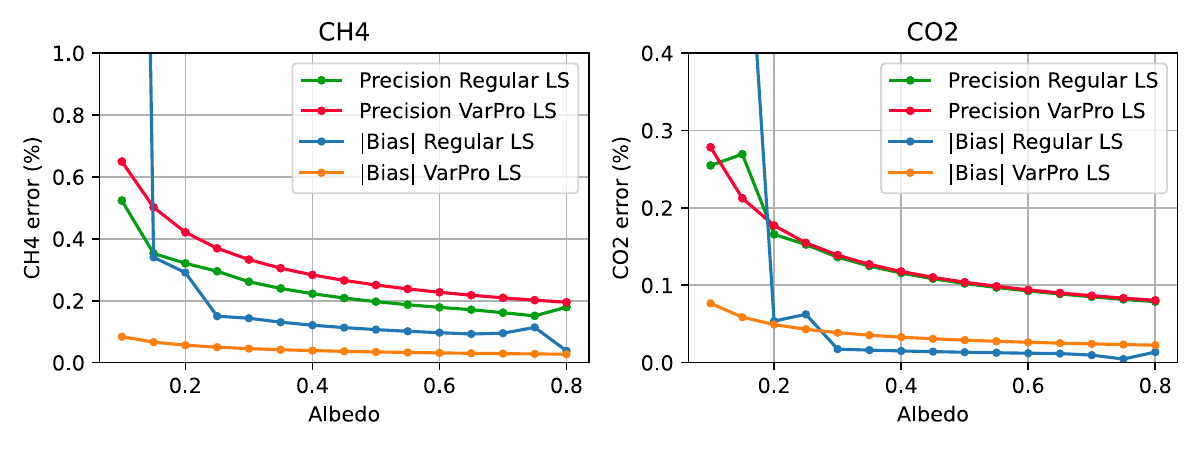}
    \caption{Comparison between the regular least-square optimization and variable projection least square optimization for the CHROMA-D configuration. The absolute bias is shown for both retrieval optimizations (blue and orange lines) together with the precision (green and red lines) }
    \label{fig:varpro}
\end{figure}

The retrieval precision, bias and total error for the CHROMA-D configuration are shown in Fig.~\ref{fig:varpro} for both the conventional nonlinear least-squares retrieval and the variable-projection retrieval introduced in Section~\ref{sec:retrieval}. The total retrieval error is defined as $
e = \sqrt{\sigma^2+b^2}$ where $\sigma$ is the standard deviation of the retrieved values and $b$ is the retrieval bias with respect to the true value. For the conventional nonlinear least-squares retrieval, the bias increases strongly towards low albedo as the measurement noise increases. Variable projection substantially reduces this effect: over the investigated albedo range, the remaining bias is approximately 5--10 times smaller than the retrieval precision. The total retrieval error is therefore predominantly determined by the measurement noise rather than by retrieval bias.

\begin{figure}[h]
    \centering
    \includegraphics[width=\linewidth]{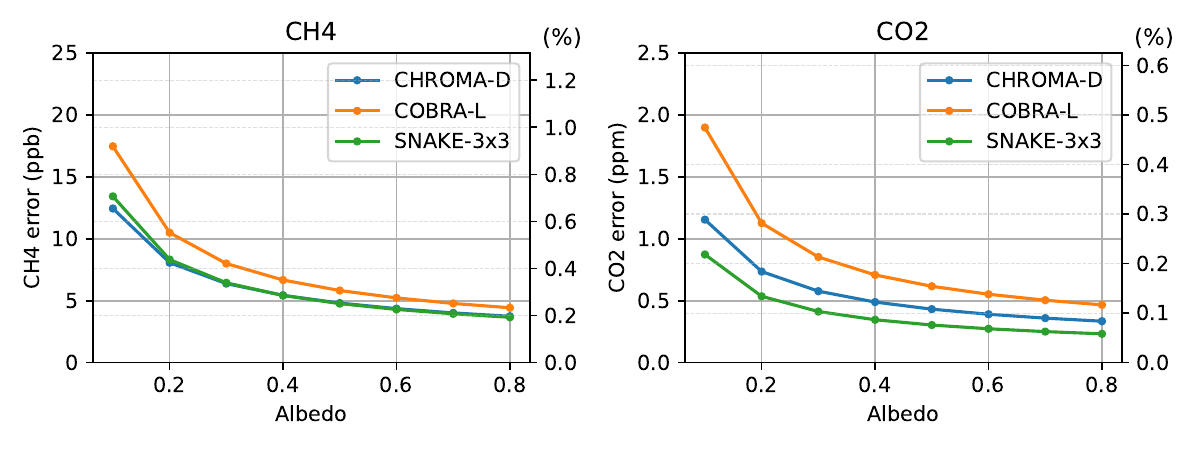}
    \caption{Performance of the three instrument configuration as function of albedo level evaluated for the specified GRE. For all these retrievals a sun zenith angle of 50\textdegree was used.}
    \label{fig:config_performance}
\end{figure}

Figure~\ref{fig:config_performance} shows the total retrieval error as a function of surface albedo for all three instrument configurations. The CHROMA-D and SNAKE $3\times3$ configurations provide the best overall performance, with the SNAKE configuration showing a small advantage for the CO$_2$ retrieval. For surface albedos of 0.2 and higher, both configurations achieve CH$_4$ retrieval errors below approximately 10~ppb and CO$_2$ retrieval errors below approximately 0.7~ppm. All simulations shown here were performed for a solar zenith angle of 50\textdegree.

These values are comparable to the reference precision requirements of the CO2M mission of 10~ppb for CH$_4$ and 0.7~ppm for CO$_2$ at an albedo of 0.2. The proposed configurations achieve this level of simulated retrieval performance with ground resolution elements of approximately 250--300~m, compared with the $2\times2$~km$^2$ spatial resolution of CO2M. The results therefore indicate that the photonic-crystal filter architecture could enable global CH$_4$ and CO$_2$ mapping at substantially finer spatial resolution while maintaining retrieval performance comparable to state-of-the-art global mapping missions. For a swath of 150 km, global coverage in 4 days can be achieved with 5 satellites. Alternatively, to decrease the revisit time or number of satellites the number of camera modules per satellite could be increased. Furthermore, the CH4/CO2 proxy method \cite{MethanAir2024} can be used  in the SWIR-1 range to correct for common light path errors. This is suitable for enhancement measurements when the proxy (for example CO2 for CH4 detection) is well constrained. However, for a full physics retrieval auxiliary information is required, which would drive for example the addition of a multi-angle spectropolarimeter \cite{rusli2021anthropogenic}. 

In addition to the retrieval performance, the proposed architecture offers an inherent reduction in spectral data volume. Each ground element is represented by the integrated signals of only ten task-optimized filters rather than by a densely sampled spectrum. With appropriate on-board temporal and spatial coadding, the detector data can therefore be substantially reduced before downlink. This is particularly attractive for small-satellite and constellation applications, where available downlink capacity can be a limiting system resource.
 
\section{Outlook: Filter manufacturing and breadboard demonstration}
\begin{figure}
    \centering
    \includegraphics[width=0.65\linewidth]{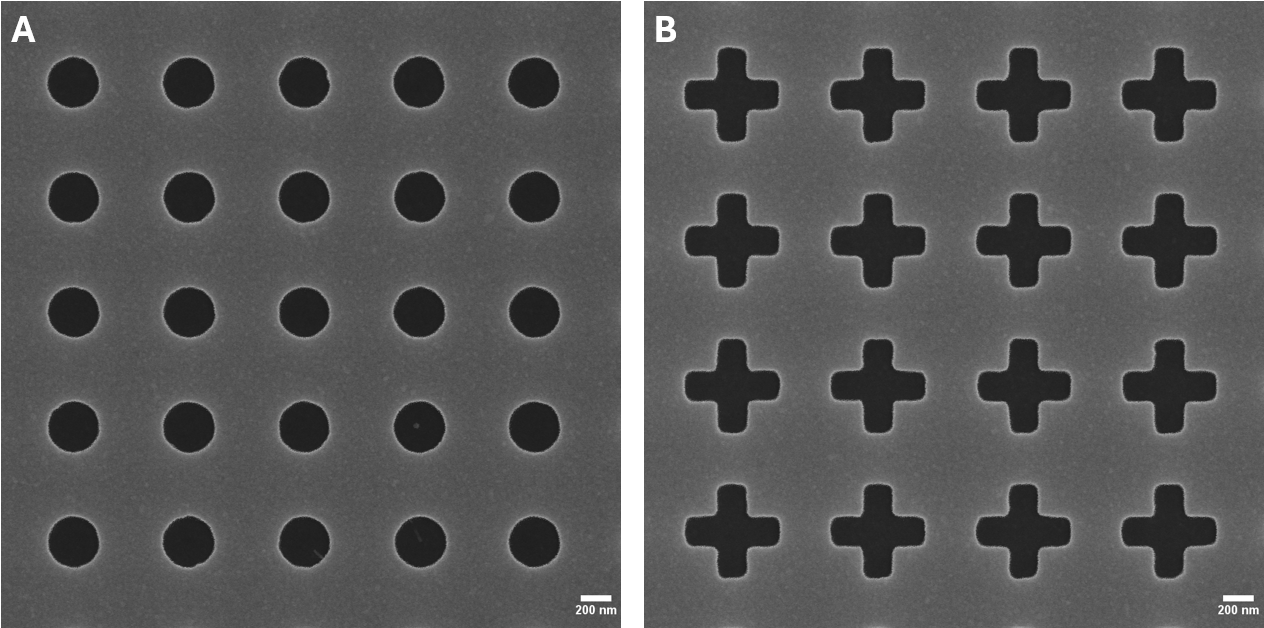}
    \caption{SEM images of silicon photonic crystals with circles (A) and crosses (B), manufactured by NIL Technology. Scale bar indicates 200~nm.}
    \label{fig:SEMimages}
\end{figure}

The presented performance in this work is based on end-to-end simulations and represents a technology projection; experimental validation of the photonic-crystal filters and integrated detector assembly is the next step. In the coming period, we will concentrate on the experimental demonstration of the filter manufacturing, testing and filter integration on a Lynred SNAKE detector. Silicon photonic crystals are currently under manufacturing with e-beam lithography at NIL Technology in Denmark. In Fig.~\ref{fig:SEMimages}, examples of Scanning Electron Microscope (SEM) images of different photonic crystal types can be seen.

The photonic crystals will be measured in free space and after integration on the detector for a range of F-numbers. %For the detector integration on TEC Lynred SNAKE detector, the lid of the sealed detector housing will be removed. 
These measurements will allow the experimentally obtained filter responses to be compared with the simulated transmission profiles used for the filter selection and performance simulations presented in this work. Following successful filter fabrication and detector integration, the photonic crystal-detector assembly can be combined with a telescope to form a compact camera module for greenhouse-gas observations. The low predicted mass and volume of such a module could additionally enable demonstration from a high-altitude platform such as Zephyr.

Although the present work focuses on CH$_4$ and CO$_2$, the same task-optimized filter approach can be adapted to other trace gases with suitable absorption features, such as CO and NH$_3$, by selecting the spectral range and photonic-crystal filters for the corresponding retrieval problem. Experimental validation of the integrated filter-detector assembly will provide the next step towards demonstrating the feasibility of this architecture for compact, high-resolution trace-gas mapping. 

\bibliography{references} % bibliography data in report.bib
\bibliographystyle{spiebib} % makes bibtex use spiebib.bst

\end{document}